# SEMI-AUTOMATED ANNOTATION OF SIGNAL EVENTS IN CLINICAL EEG DATA

*S. Yang, S. López, M. Golmohammadi, I. Obeid and J. Picone*

Neural Engineering Data Consortium, Temple University, Philadelphia, Pennsylvania, USA
{scott.yang, silvia.lopez, meysam, obeid, picone}@temple.edu

*Abstract*— To be effective, state of the art machine learning technology needs large amounts of annotated data. There are numerous compelling applications in healthcare that can benefit from high performance automated decision support systems provided by deep learning technology, but they lack the comprehensive data resources required to apply sophisticated machine learning models. Further, for economic reasons, it is very difficult to justify the creation of large annotated corpora for these applications. Hence, automated annotation techniques become increasingly important.

In this study, we investigated the effectiveness of using an active learning algorithm to automatically annotate a large EEG corpus. The algorithm is designed to annotate six types of EEG events. Two model training schemes, namely threshold-based and volume-based, are evaluated. In the threshold-based scheme the threshold of confidence scores is optimized in the initial training iteration, whereas for the volume-based scheme only a certain amount of data is preserved after each iteration. Recognition performance is improved 2% absolute and the system is capable of automatically annotating previously unlabeled data. Given that the interpretation of clinical EEG data is an exceedingly difficult task, this study provides some evidence that the proposed method is a viable alternative to expensive manual annotation.

## I. INTRODUCTION

An electroencephalogram (EEG) is a nonintrusive clinical tool that has been widely used by neurologists and clinicians to diagnose brain-related illnesses such as epilepsy and seizures [1]. However, manual annotation of clinical data is both time-consuming and expensive, requiring well-trained board-certified neurologists that are in short supply. The interpretation of the EEG signals itself is quite subtle and challenging. For example, in [2] the inter-rater agreement on identification of electrographic seizures had a kappa statistic of 0.58. For periodic discharges, the kappa statistic was 0.38. On more subtle events, such as spikes and sharp waves, or the onset of non-convulsive seizures, the degree of inter-rater agreement is even lower.

Further, it is virtually impossible to generate the vast amount of EEG data needed to train advanced deep learning systems by manual annotation. Such systems often only show their effectiveness when there is a large amount of data. Therefore, an automatic annotation process is highly desirable because it is a cost-effective way to generate large amounts of data. It would allow more niche problems in the bioengineering field to be addressed by powerful machine learning technology.

Fortunately, machine learning techniques make it possible to automatically annotate big data. Semi-supervised learning methods such as self-training [3] and active learning [4] are attractive in this scenario. Both approaches require a small amount of transcribed data and iteratively re-train models for improved classification. However, compared with a self-training algorithm, conventional active learning still requires constant cooperation from expert(s) to interactively annotate the most informative (least confident) data. Self-training, one of several semi-supervised approaches to active learning [5], selects only the most confident features for automatic model re-training. In this study, we developed a self-training approach to iteratively annotate a large EEG corpus. Figure 1 shows a generic flow diagram of the self-training process.

Self-training has been employed successfully in some applications, such as face recognition [6], word classification [7], object detection [8] and gait recognition [9]. Li et al. [10] applied self-training to an EEG-based brain computer interface spelling system. Their baseline system was trained by a support vector machine (SVM) [11]. Panicker et al. [12] reported an EEG recognition system using a two-classifier co-training approach to improve the classification of P300 visual evoked potentials. Both of the studies claimed substantial recognition improvements. However, in clinical applications where large amounts of EEG data exists, there has been no positive result reported in the literature.

## II. METHODOLOGY

A standard pattern recognition system usually contains two major modules: feature extraction and pattern classification. In this study, the raw EEG signals were first segmented into 1-second epochs and processed through a standard feature extraction process described in detail in [14]. The baseline system, also described in [14], is designed to identify six classes of events: (1) spike and/or sharp waves (SPSW), (2) periodic lateralized epileptiform discharges (PLED), (3) generalized periodic epileptiform discharges (GPED), (4) artifacts (ARTF), (5) eye movements (EYEM) and (6) background (BCKG). The first three events are associated with signal events and are of clinical interest. The last three events are associated with the background channel and are simply used to improve classification performance.

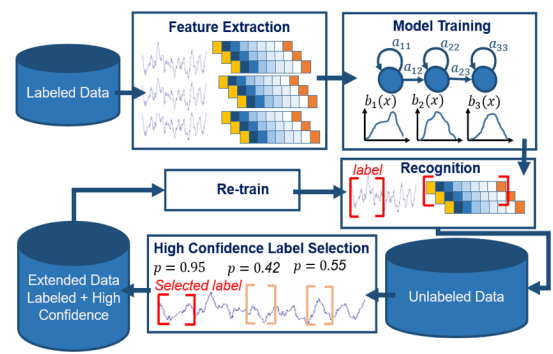

Figure 1. A generic self-training approach

In this work, self-training is employed to automatically annotate the unlabeled data in the Temple University Hospital EEG Corpus (TUH EEG) [15]. TUH EEG is the world's largest publicly available database of clinical EEG data, comprising more than 30,000 EEG records and over 16,000 patients. Though we have EEG reports for all of the data, the signal data is not fully annotated. In order to make use of this data for machine learning, particularly signal event classification, we need annotated data. Self-training becomes an attractive option. Self-training usually employs a small amount of labeled data to train the classifier, and then uses the trained classifier to predict (decode) the unlabeled data. The effectiveness of self-training on EEG data has been mixed [10] [12]. For example, Panicker et al. [12] found that the addition of unlabeled/auto-labelled data can only improve the system performance to certain limit, the addition of more unlabeled data will degrade the performance: the degradations were spotted during the analysis of data from all the five subjects who participated the experiments. This effect has been also reported by Cohen et al. [13] using different datasets for semi-supervised learning on human-computer interactions. Our experience has been that these results do not extend to the large amounts of clinical data described here, and hence the motivation for this work.

The proposed annotation system begins with a very small amount of manually annotated EEG data. We used a bootstrapping process to create this seed data. We initially had an expert transcribe a very small amount of data (100 10-second segments) for the three signal classes (SPSW, GPED and PLED). The remaining classes (BCKG, ARTF and EYEM) are relatively easy to identify, so we were able to train a team of undergraduates to annotate these classes.

We trained a background model on a large portion of the remaining data since the vast majority of the data is background. We then built a pilot set of models and used these to locate events of interest. These events were then manually reviewed. We iterated this process until we had a decent set of baseline models. The performance of the baseline system was 65.0% for six-way classification. Each model was trained by manually labelled data from 290 patients, which contained 3.1 hours (11,253 seconds) of PLED, 1.7 hours (6,161 seconds) of GPED, 10 minutes (643 seconds) of SPSW, 15 hours (53,726 seconds) of BCKG, 3.1 hours (11,053 seconds) of ARTF segments and 18 minutes (1,070 seconds) of EYEM.

Once the baseline system was stabilized, we then attempted to use these models and our self-training method to label the remaining data. The control flow for this process is shown in Figure 1 and can be described as follows:

1. Development of the initial machine learning algorithm to train the six classes of models;
2. Decode (predict) the unlabeled database using the trained models.
3. Extract confidence scores based on posteriors from each class and select the corresponding events that have high confidence labels.
4. Add the selected high-confidence events to the training pool and re-train the models.
5. Evaluate the updated models using a standard open-set evaluation.
6. Use the updated models to decode the rest of the corpus.
7. Repeat steps 2 to 6 until the entire database is labeled with high confidence.

The proposed algorithm selects the best candidates for a given class for re-training of the original models. The original Hidden Markov Model (HMM) for training contained eight mixtures with five states [14]. As long as the quality of the selected features is decent, with additional iterations of the re-training process, we expect the recognition performance to continuously increase. Of course, the optimal way to identify those patterns that will force the system to improve itself is the art of semi-supervised learning approaches.

The proposed self-training method does not need human intervention by an expert between iterations. This is what separates our approach from many published results obtained from conventional active learning. The algorithm selects the most highly-ranked events and adds them to the existing training set after each re-training round. Since we do not have truth markings for the data, we used an event-specific likelihood threshold to determine the most suitable patterns to be added to the training set. It is worth mentioning that confidence scores measured in the experiments were the normalized probability densities derived from the standard forward-backward algorithm [16]. Classification performance was evaluated on the manually-labeled evaluation set to determine if performance was improving performance and/or converging. Each of the six models used different thresholds which were adjusted automatically after each iteration.

Clinical data is significantly more challenging than data collected under controlled conditions because there are many more artifacts and channel variations. These artifacts pose serious problems for semi-supervised classification schemes such as that used in active learning. For example, a movement artifact can often be confused with an SPSW event, making it a promising, but misleading candidate for active learning. Self-training algorithms tend to be sensitive to the outliers, which can lead to an entire cluster converging to the wrong label [3]. In the following section we present some pilot results of these approaches on TUH EEG.

## III. EXPERIMENTAL ANALYSIS

A series of experiments were conducted to investigate using the proposed algorithm to automatically annotate TUH EEG. Since the SPSW event is fairly rare and difficult to annotate, it is one type of event that we expect to benefit significantly from this process. Therefore, some of the pilot studies used the data of this class to analyze and optimize parameters.

*3.1 Effectiveness of re-training for each class*

We investigated the effectiveness of self-training in the first iteration for all the six classes separately. For each class, about half of the events in the original training set were selected for re-training. For example, the original training set contained 6,161 GPED epochs, after the first re-training and decoding process, about 3,000 top-ranked GPED epochs were selected to augment the training set.

The main purpose of this investigation is to analyze recognition performance of each class after self-training was applied. Table 1 shows the sensitivity performance of six classes. Sensitivity improved about 8% absolute for the SPSW class after the first round of re-training. The sensitivity for the GPED and PLED classes also improved around 4% and 6% absolute respectively. However, the self-training algorithm seems to be less effective for EYEM. Further, the BCKG and ARTF classes also suffered a small degradation in performance. Since these are background events, the need for active learning for these classes is not as critical since there is a large amount of background data available in the corpus. A detailed exploration on the volume of data preserved for re-training is presented in the next section.

Table 1. Sensitivity after the first iteration of self-training

| Class | Before | After |
|-------|--------|-------|
| GPED  | 52.8%  | 56.5% |
| PLED  | 54.2%  | 60.4% |
| SPSW  | 41.6%  | 49.6% |
| EYEM  | 81.8%  | 82.1% |
| BCKG  | 72.1%  | 71.2% |
| ARTF  | 41.2%  | 39.1% |

*3.2 The number of preseved events for re-training*

As was mentioned in Section II, a key step in the self-training process is to select decoded epochs with high confidence. In this pilot study, we investigated the impact of different rankings of epochs. We focused on the SPSW class for this parameter analysis. Figure 2 shows the trend in recognition performance when we reduce the number of included SPSW events during re-training. In this analysis, we controlled the amount of highly ranked epochs: by reducing the preserved features for re-training, an increasing of the recognition performance was observed.

We began by augmenting the training set with the top 10% of the decoded features. As we tightened the inclusion thresholds, recognition performance increased. Given that the original training set contained 643 SPSW epochs, we added roughly half of the top-ranked newly predicted SPSW epochs (320 epochs) into the new training set (the fourth column in Figure 2). The new model was able to achieve a sensitivity of 65.1%. The purpose of this analysis was to find an effective scheme for adding decoded events to the original training set during the re-training iterations.

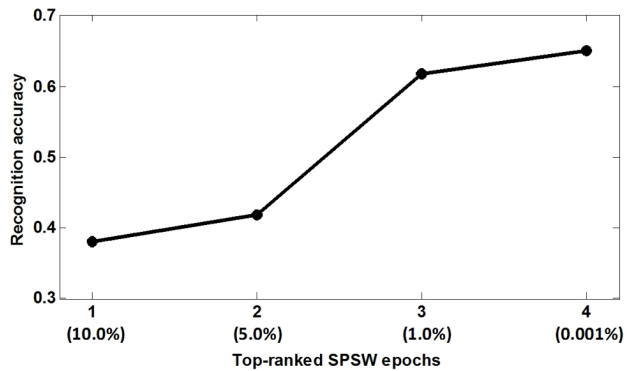

Figure 2. Preserved SPSW epochs for re-training

We also investigated the number of selected epochs as a parameter per iteration given a threshold. In this analysis, only 100 unlabeled EEG files were employed to speed up the analysis. Figure 3 shows the increase in the number of selected SPSW epochs as a function of the re-training iteration. In the first iteration only 30 events with the highest confidence scores (a threshold above 390 for the log likelihood) were added into the original training set. In the second iteration 516 events were added.

We observe that as we increased the number of iterations, as many as 28,307 SPSW epochs were preserved and recognition performance increased. However, recognition accuracy begins to drop from the third iteration. Both Figure 2 and Figure 3 indicate that considering the volume of unlabeled data, the size of the original labeled data is more influential. The last column of Figure 2 and the second column of Figure 3 suggest that it is better to grow the size of the re-training database by approximately half in each re-training iteration.

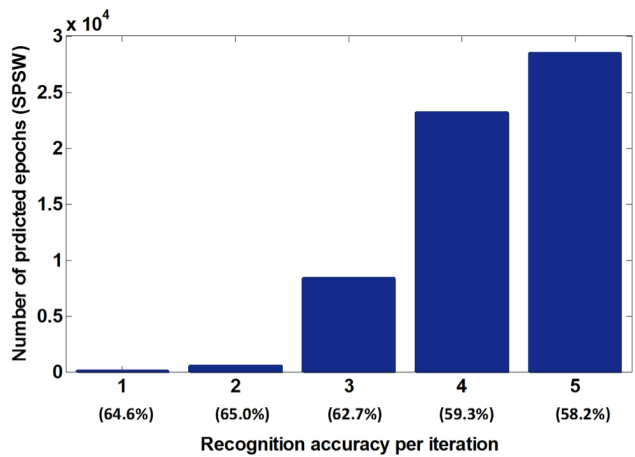

Figure 3. Impact of the number of preserved events on performance

### 3.3 Threshold analysis

Intuitively, we expect an increase in the re-training threshold will eliminate low-quality events. We conducted a series of experiments with different thresholds: based on the observations from Section 3.1, we optimized the threshold for SPSW in the initial iteration. We chose these thresholds based on the results from Section 3.2.

Figure 4 clearly shows that the best performance was achieved while the threshold was set to 355, which resulted in adding new 2,992 SPSW decoded events to the original training set. Interestingly, recognition performance begins to decrease after the peak, even though theoretically the selected events with the higher thresholds should allow better re-training of the models. When the threshold was set to 375, only 24 new SPSW events were selected. A small amount of additional high confidence data does not have a significant impact on model re-training. Note that despite this degradation in recognition performance, the performance is still slightly better than the baseline accuracy.

### 3.4 Comparision of the two experimental schemes

In previous sections we investigated a number of factors which affect the proposed algorithm when it was applied to TUH EEG. In this section, two data selection schemes were analyzed: 1) a volume-based scheme: after optimization in the first iteration, in subsequent iterations roughly half of the previously selected epochs were added; and 2) a threshold-based scheme: the threshold is optimized in the first iteration and this threshold is unchanged in the following iterations. Scheme 1 (S1) may lead to an increase in the threshold after each new iteration, whereas scheme 2 (S2) keeps the optimized threshold fixed.

Figure 4 shows the performance of the two schemes. Seven iterations were conducted to demonstrate trends in performance for the proposed self-training method. These two schemes were demonstrated using only data from the SPSW class. The first iteration was optimized based on the findings in Section 3.2 (Figure 2 and Figure 3). Using the parameters optimized in Section 3.3 (Figure 4) from S1, the overall performance increased about 2% absolute. S2 kept the initial threshold unchanged and the performance continuously degraded. This suggests that the threshold optimized in the initial iteration is no longer optimal for subsequent iterations.

## IV. SUMMARY

In this paper we investigated using self-training as a semi-supervised algorithm to automatically annotate the TUH EEG Corpus. The proposed method is based on a limited amount of manually annotated data and updates the training models by iteratively re-training and re-decoding. This pilot study shows after a few iterations we are able to not only label more EEG signals (about 30,000 new SPSW labels) but also improve recognition accuracy about 2%. Though the scale of these improvements is small, this is the first time this approach has worked on large amounts of clinical EEG data. For example, in [10] and [12] the authors reported EEG-based recognition systems tested by pools of 10 and 5 subjects, respectively. Further improvements can be made by employing new techniques that integrate a variety of learning algorithms for active learning. It is also worthwhile to explore other auto-labelling algorithms such as co-training, which may potentially improve the recognition performance.

## ACKNOWLEDGMENTS

This material is based in part upon work supported by the National Science Foundation (NSF) under Grant No. IIP-1622765. Any opinions, findings, and conclusions or recommendations expressed in this material are those of the author(s) and do not necessarily reflect the views of the NSF. Research reported in this publication was also supported by the National Institutes of Health under Award Number U01HG008468. The TUH EEG Corpus effort was funded by (1) the Defense Advanced Research Projects Agency MTO under the auspices of Dr. Doug Weber through the Contract No. D13AP00065, (2) Temple University's College of Engineering and (3) Temple University's Office of the Senior Vice-Provost for Research.

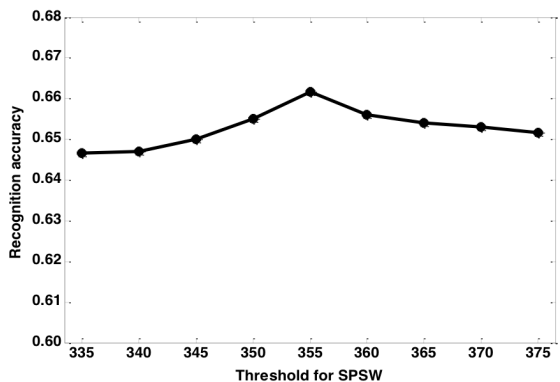

Figure 5. Threshold optimization in the initial iteration

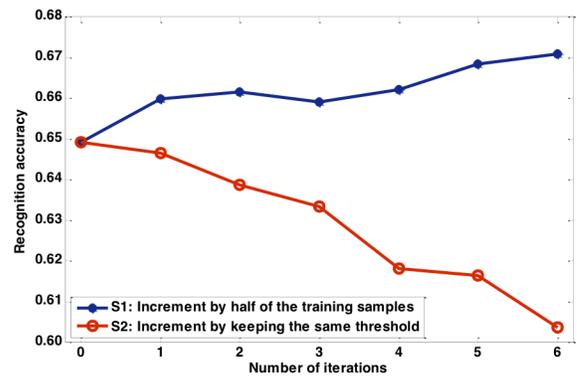

Figure 4. Comparison of two schemes